\documentclass[conference]{IEEEtran}
\usepackage{graphicx}
\begin{document}
%
\title{Robotic Control for Cognitive UWB Radar}

\author{\IEEEauthorblockN{Stefan Br\"uggenwirth}
\IEEEauthorblockA{Fraunhofer Institute for High Frequency \\Physics and Radar Techniques (FHR)\\Fraunhoferstr. 20, 53343 Wachtberg, Germany\\Email: stefan.brueggenwirth@fhr.fraunhofer.de}
\and
\IEEEauthorblockN{Fernando Rial}
\IEEEauthorblockA{Fraunhofer Institute for High Frequency \\Physics and Radar Techniques (FHR)\\Fraunhoferstr. 20, 53343 Wachtberg, Germany\\Email: fernando.rial@fhr.fraunhofer.de}
}


%


\maketitle

\begin{abstract}
In the article, we describe a trajectory planning problem for a 6-DOF robotic manipulator arm that carries an ultra-wideband (UWB) radar sensor with synthetic aperture (SAR). The resolution depends on the trajectory and velocity profile of the sensor head. The constraints can be modelled as an optimization problem to obtain a feasible, collision-free target trajectory of the end-effector of the manipulator arm in Cartesian coordinates that minimizes observation time. For 3D-reconstruction, the target is observed in multiple height slices. For Through-the-Wall radar the sensor can be operated in sliding mode for scanning larger areas. For IED inspection the spot-light mode is preferred, constantly pointing the antennas towards the target to obtain maximum azimuth resolution.
\end{abstract}


%
\IEEEpeerreviewmaketitle

\section{Introduction}

\begin{figure*}[ht!]
	\centering
	\includegraphics[width=0.75\textwidth]{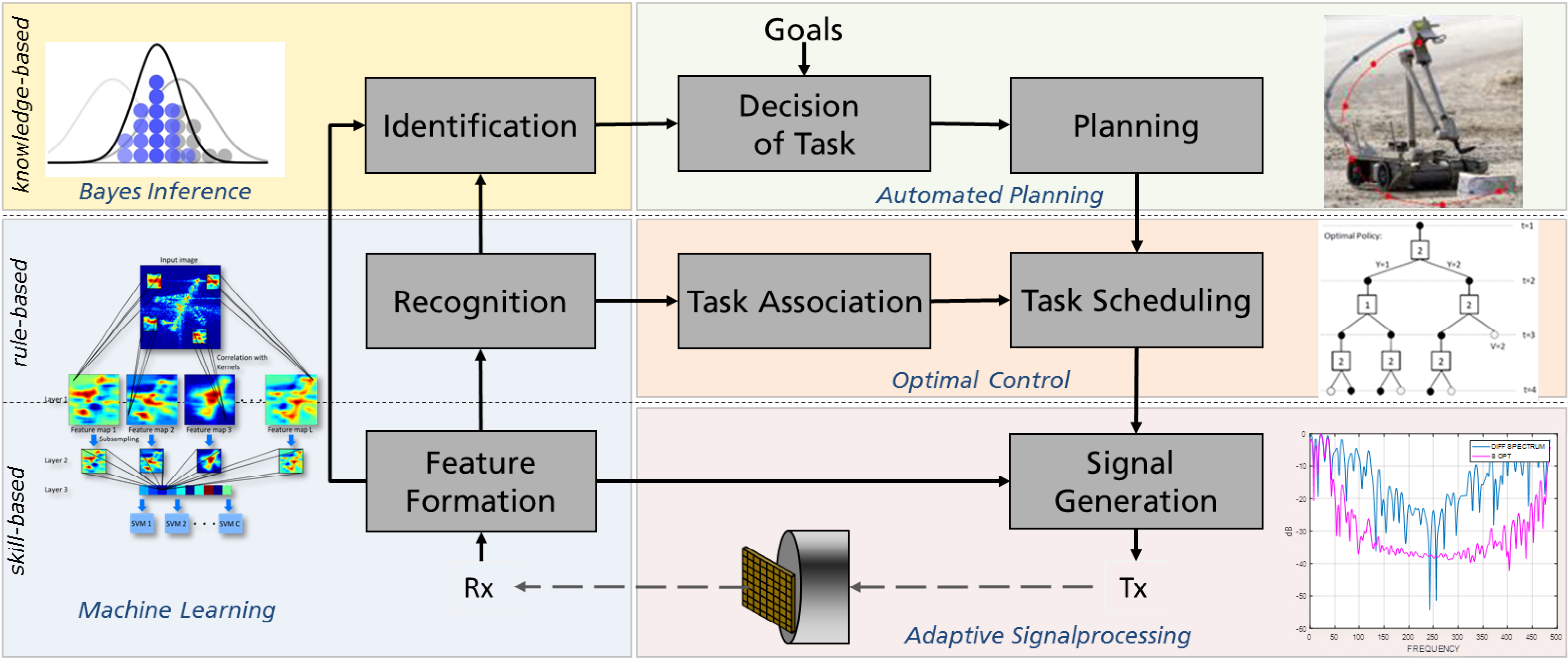}
	\caption{Three-layer model of a cognitive radar architecture according to \cite{rasmussen1983skills}}
	\label{fig:dreiEbenenModell}
\end{figure*}

The concept of a cognitive radar \cite{haykin2012cognitive,guerci2010cognitive} architecture at Fraunhofer FHR \cite{ender2015cognitive} is based on the three-layer-model by Rasmussen \cite{rasmussen1983skills} as shown in Fig.\ref{fig:dreiEbenenModell}. 

The \textit{skill based} behavior represents the basic signal generation and processing capabilities of the system. This provides the subsymbolic, continuous stream of input-data to the architecture. The \textit{rule-based} abstraction layer applies machine-learning methods to recognize certain pre-stored cues in the perceived scene. Upon match, the architecture will reactively execute a prestored procedure on its actuators or modify its sensor parameter settings. The \textit{knowledge-based} abstraction layer provides long-term deliberation and goal-based planning. 

In this article, we will focus on the generation of constraints for trajectory planning for a robotic arm carrying a UWB radar sensor. All three layers need to interact to compute the trajectory according to the scene and execute it by commanding the sensor and the manipulator joints and finally form an image.

Ultrawide-band 3D-SAR imaging using robotic arms have been demonstrated using milllimeter-Wave FMCW-Radars \cite{yang2012fmcw, lang20153d, wang2016effects}. In this 
article, we concentrate on a chirped radar with a lower center frequency of 8 GHz, yielding better material penetration properties.

\section{Robotic UWB sensor setup} 

\begin{figure}[ht!]
	\centering
	\includegraphics[width=0.4\textwidth]{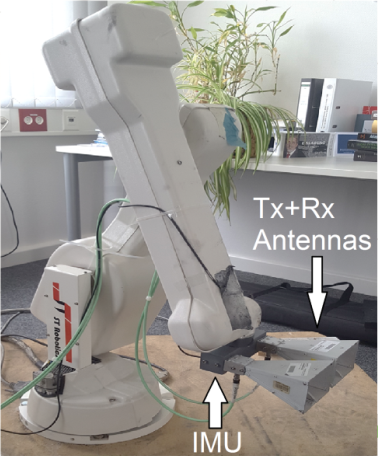}
	\caption{Transmit (Tx) and receive (Rx) antenna of UWB radar sensor mounted to robotic manipulator arm. Photo also shows IMU position.}
	\label{fig:CSPTrajektorie}
\end{figure}

The system under consideration consists of a robotic manipulator arm that carries a UWB sensor able to work in synthetic aperture radar (SAR) mode. We used a stock ST Robotics R17 6-DOF industrial type manipulator arm shown in (Fig. \ref{fig:CSPTrajektorie}).

High resolution imaging can be only achieved with an even higher precision positioning. The 3D-trajectory of the sensor needs to be measured and synchronized with the sensor data. For that purpose, accelerometers and gyroscopes from an attached inertial measurement unit (IMU) are used. The IMU drift is additionally stabilized using the hardware readout of optical encoders of the robot arm joints controlled by step-motors.

\begin{figure}[ht!]
	\centering
	\includegraphics[width=0.45\textwidth]{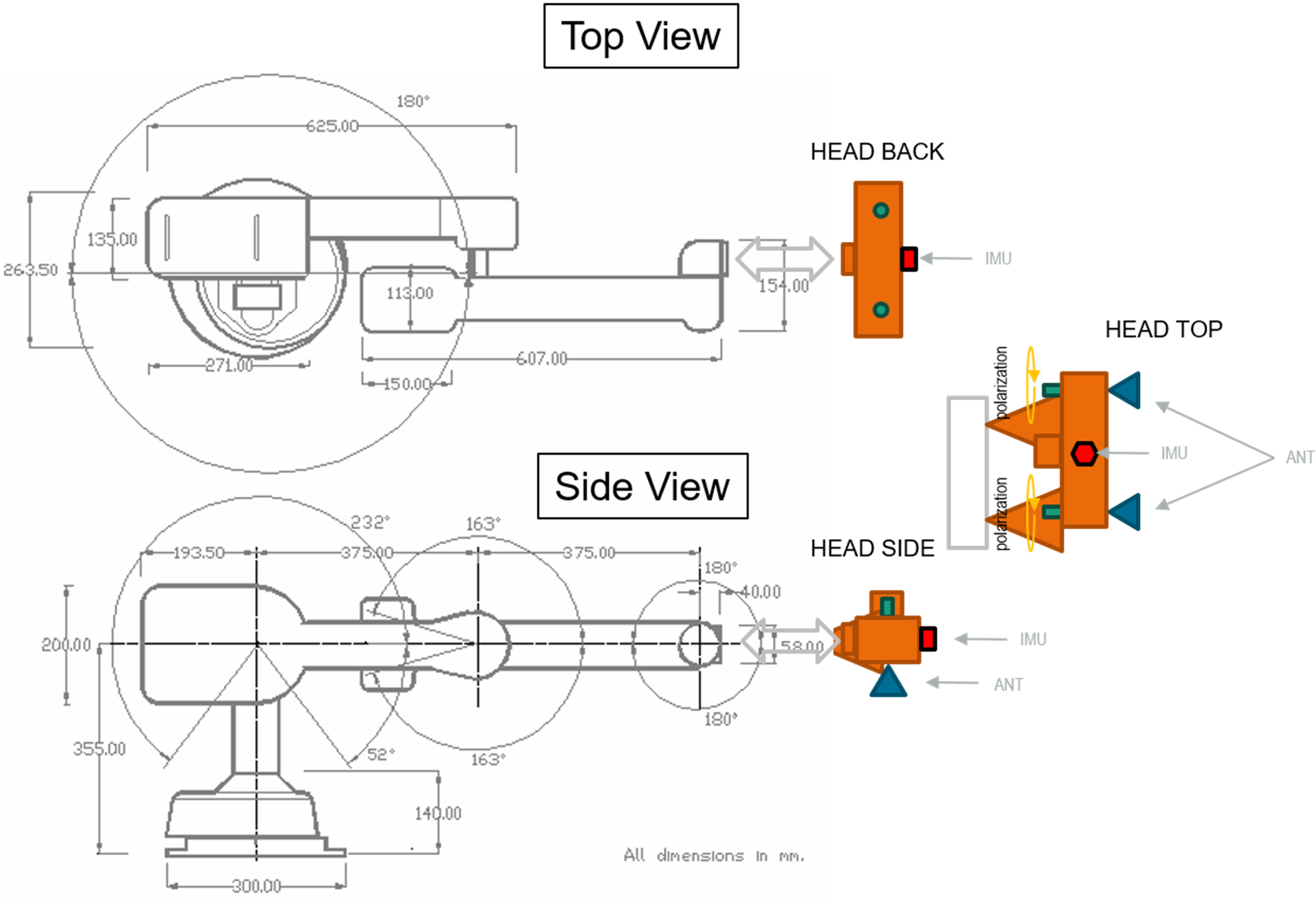}
	\caption{Schematic of the robot arm and polarimetric antenna mounting.}
	\label{fig:RobotScheme}
\end{figure}

The sensor has one transmitter and one receiver in a typical common offset arrangement as shown in Fig. \ref{fig:RobotScheme}. The horn type antennas can be rotated to exploit polarization diversity. 

The spatial resolution and processing gain that the system can achieve ultimately depend on the trajectory and velocity profile of the sensor head. The constraints can be modelled as an optimization problem to obtain a feasible, collision-free trajectory of the end-effector of the manipulator arm in Cartesian coordinates that minimizes observation time.

\section{Sensor Characteristics and Trajectory Constraints} 
The chirped radar sensor uses a selectable center frequency from 3 to 8GHz and 4GHz of bandwidth B, resulting in $\delta_{r}=3.75cm$ of range resolution given by:
\begin{equation}
\delta_{r} \simeq \frac{c}{2B}
\end{equation}

where c denotes the speed of light in vacuum.

The center frequency can be tuned according to a particular target or propagation environment (ground penetration, through-the-wall imaging, IED inspection, etc.). The sensor is able to operate in stripmap or spotlight SAR modes using linear trajectories. Several parallel trajectories can be combined for 3D imaging. The mobility of the arm could be further exploited to generate non-linear trajectories around a target to obtain a more accurate 3D reconstruction.

\begin{figure}[ht!]
	\centering
	\includegraphics[width=0.4\textwidth]{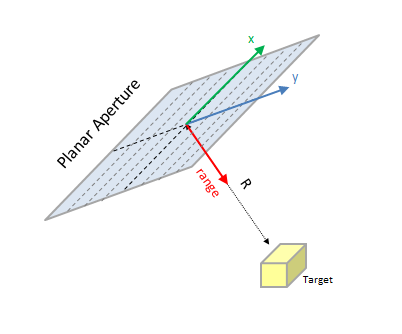}
	\caption{Planar aperture using linear trajectories.}
	\label{fig:PlanarAperture}
\end{figure}

Considering the case of planar acquisition geometries working in stripmap mode (as in Fig. \ref{fig:PlanarAperture}) the cross-range resolutions $\delta_{x,y}$ can be generally obtained as:

\begin{equation}
\delta_{x} \simeq R\frac{\lambda_{c}}{2l_{x}}
\end{equation}
\begin{equation}
\delta_{y} \simeq R\frac{\lambda_{c}}{2l_{y}}
\end{equation}

where R is the distance to the target,Fig.5  $\lambda_{c}$ is the wavelength at the center frequency and $l_{x}$ and $l_{y}$ are the lengths of the 2D synthetic aperture (during which the target is in the antenna beam).

Therefore, in order to obtain a similar resolution in range and in cross-range, the trajectory planning must aim to create at least an aperture of 0.5 to 1.3 times the distance R to the target in both directions, depending on the center frequency used by the system (8-3GHz respectively).

Assuming the target is continuously sliding through the antenna beam, another way to express the cross-range resolution relies on the knowledge of the antenna footprint:

\begin{equation}
\delta_{x} \simeq \frac{\lambda_{c}}{2\Theta_{x}}
\end{equation}
\begin{equation}
\delta_{y} \simeq \frac{\lambda_{c}}{2\Theta_{y}}
\end{equation}

where $\Theta_{x}, \Theta_{y}  $ are the antenna beam angles in radians in the x and y directions.

 In the particular case of a sensor setup with a central frequency of 8GHz and antenna beamwidth of both $\Theta_{x}, \Theta_{y} = 20\deg$, the expected resolution will be about 5.4cm. In practice, data is windowed to reduce sidelobes in the imaging and as a result the actual resolutions are usually poorer than those given by the formulas. 

\begin{figure}[ht!]
	\centering
	\includegraphics[width=0.4\textwidth]{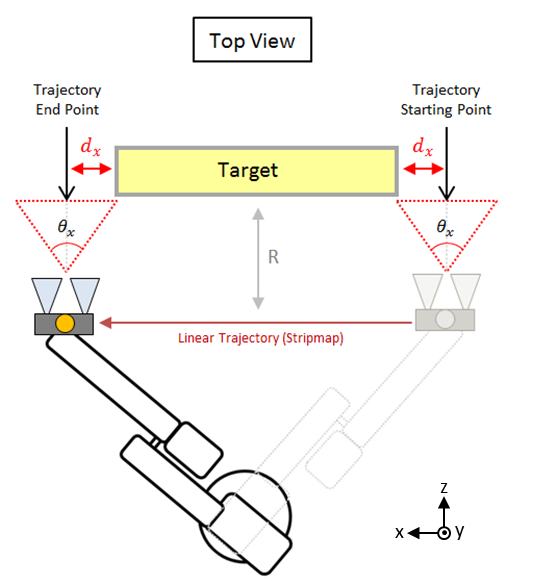}
	\caption{$x$ and $z$ direction of the scanning area.}
	\label{fig:ScanAreaTop}
\end{figure}

Two other important parameters to be taken into account are the optimal size of the scanning area and the sampling requirements. For planar acquisition geometries working in stripmap mode, to obtain full resolution imaging of the total area of interest an additional half beam aperture must be added in both dimensions as shown in Fig. \ref{fig:ScanAreaTop}.

This additional distance can be calculated with the expressions
\begin{equation}
d_{x} = R\tan{\frac{\Theta_{x}}{2}}
\end{equation}
\begin{equation}
d_{y} = R\tan{\frac{\Theta_{y}}{2}}
\end{equation}

that clearly depend on the distance between the sensor and the object of interest. 

In the particular case of an object of dimensions $D_{x} = 0.5m$  and $D_{y} = 0.3m$ located at $R = 0.5m$ from the sensor, with a central frequency of 8 GHz and antenna beamwidth of both $\Theta_{x}, \Theta_{y} = 20\deg$, the scanning trajectory should be extended by about 9 cm in both dimensions. In the following, we will use $L_{x}=0.8m$ and $L_{y}=0.5m$ as physical linear displacements with constant velocity in the $x,y$-directions to satisfy the requirements for the scanning area in both directions.

Another important parameter is related with the sampling requirements of a particular acquisition. The measurement positions in the synthetic radar aperture require a minimum spacing in order to sample adequately the phase history associated with all the scatterers.  If the distance between measurements is too large, the Nyquist criterion is not fulfilled and artifacts may appear in the reconstructed image. 

Assuming that the targeted area is confined within a rectangular box of dimensions $D_{x}, D_{y}$ and that $L_{x}$ and $L_{y}$ are the lengths of the aperture, the required sampling spacing $\Delta_{x}, \Delta_{y}$ in the measurement to satisfy the Nyquist criterion is given by \cite{fortuny2002novel}:   

\begin{equation}
\Delta_{x} \leq \frac{\lambda_{min}}{2}\cdot\frac{\sqrt{\frac{(L_{x}+D_{x})^{2}}{4} + R^{2}}}{L_{x}+D_{x}}
\end{equation}
\begin{equation}
\Delta_{y} \leq \frac{\lambda_{min}}{2}\cdot\frac{\sqrt{\frac{(L_{y}+D_{y})^{2}}{4} + R^{2}}}{L_{y}+D_{y}}
\end{equation}

where $\lambda_{min}$ is the wavelength at the maximum working frequency. In our example, the sensor-head must not move more than $\Delta_{x}=0.0095m$ per pulse. Since the pulse repetition frequency (PRF) of our system is fixed to 12Hz, the resulting maximum velocity $v_{max} = \Delta_{x} \cdot PRF = 0.11m/s$.

Here it must be considered also that signal propagation in dielectric materials (ground, wall) will shrink the wavelengths and then sampling requirements become even more stringent \cite{grasmueck2005full}. A previous estimation of the dielectric permittivity of the propagation media may further optimize the acquisition geometry.    

\section{Trajectory generation}

\begin{figure}[h!]
	\centering
	\includegraphics[width=0.45\textwidth]{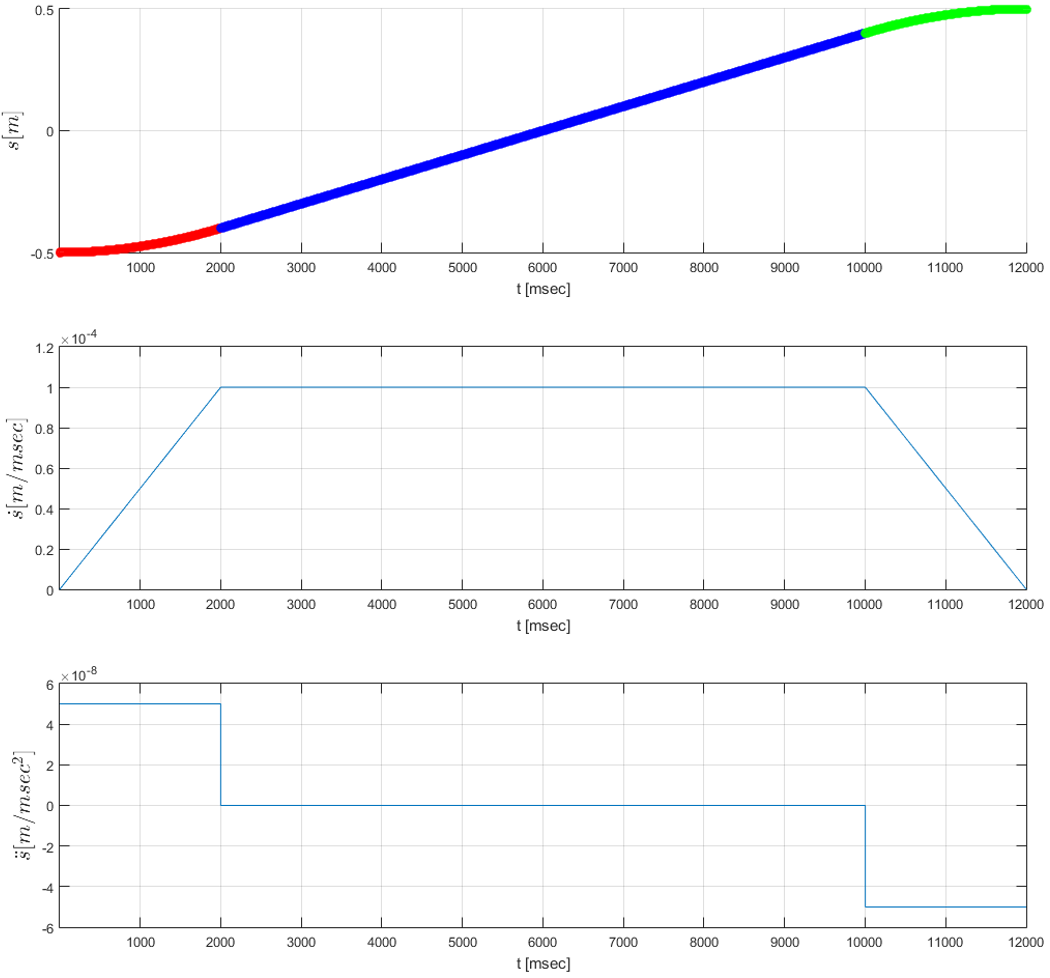}
	\caption{Trapezoidal velocity profile for linear trajectory of the sensor head.}
	\label{fig:velocityProfile}
\end{figure}

In order to maximize the acquisition speed, we selected a trapezoidal velocity \cite{corke2011} profile shown in Fig. \ref{fig:velocityProfile}. To slightly oversample, we select a constant target velocity of $v = 0.1m/s < v_{max}$. The trajectory accelerates for the first 2000msec, then reaches constant acquisition speed $v$ for $L_x=0.8m$ and then decelerates again.

\begin{figure}[h!]
	\centering
	\includegraphics[width=0.45\textwidth]{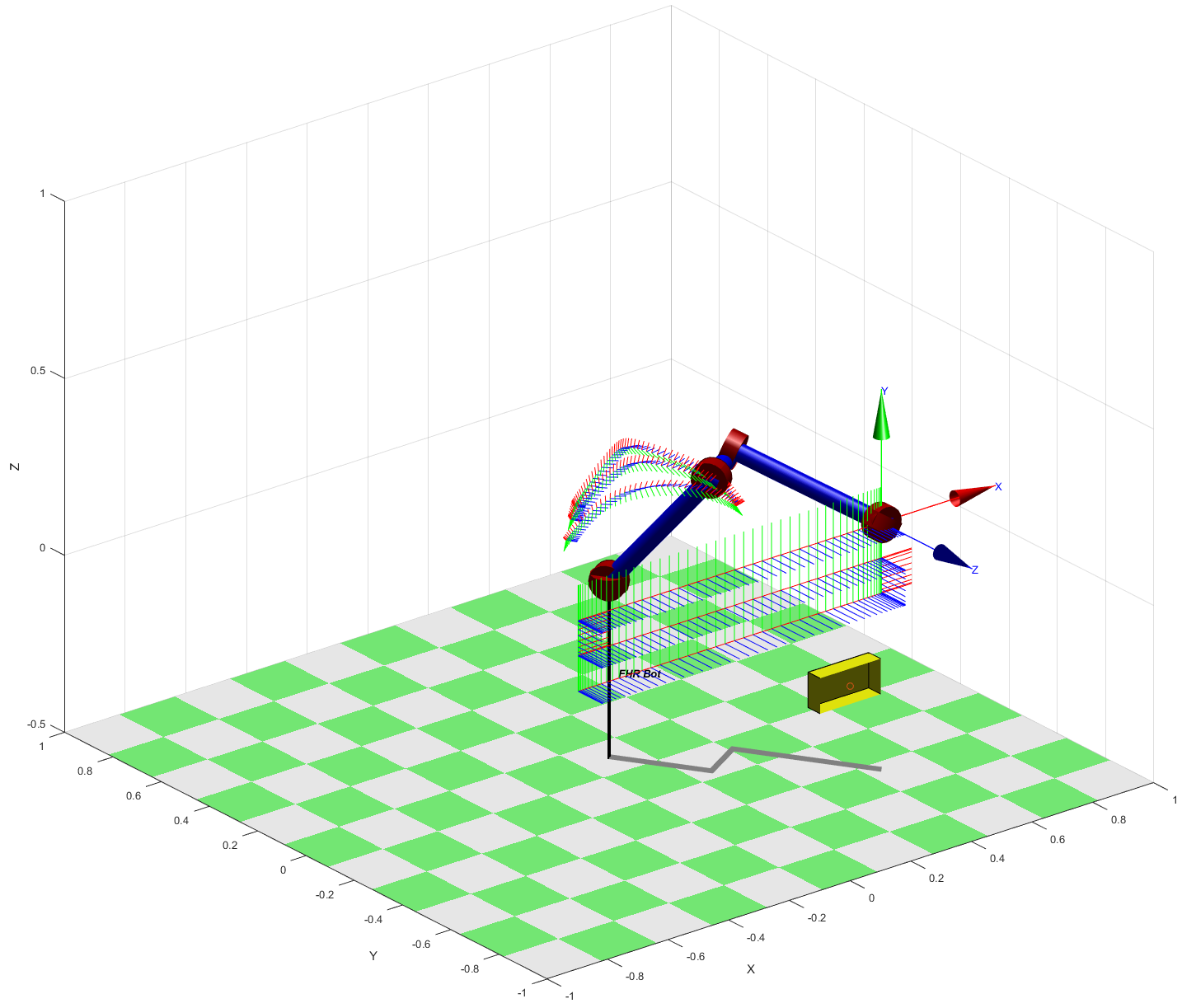}
	\caption{Resulting inverse kinematics of 6DOF robot joint for linear SAR trajectory.}
	\label{fig:6dof}
\end{figure}

The linear trajectory can be used to form a SAR image in stripmap mode for each height slice. As shown in Fig. \ref{fig:6dof}, the pose of the robot end-effector follows a meander-shaped pattern composed of 
linear segments for the different height slices. For a six degree of freedom robot, the inverse kinematics can easily be computed as illustrated in the figure for joints 3 and 6. The ST Robotics controller already 
contains routines to command the pose of the end-effector in Cartesian coordinates with the desired trapezoidal velocity profile.

\section{Experimental results}
\begin{figure}[h!]
	\centering
	\includegraphics[width=0.35\textwidth]{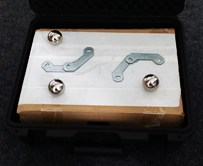}
	\caption{Photo of the objects inside the suitcase.}
	\label{fig:SuitecaseResultsPhoto}
\end{figure}

\begin{figure}[h!]
	\centering
	\includegraphics[width=0.5\textwidth]{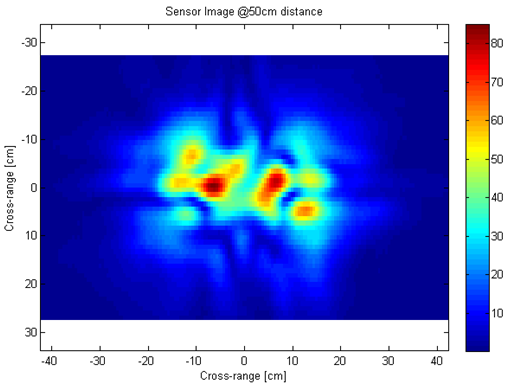}
	\caption{SAR image of the objects using the robot arm (2D slice in range).}
	\label{fig:SuitecaseResultsTop}
\end{figure}

Fig. \ref{fig:SuitecaseResultsPhoto} shows an example of an image obtained with the robot arm using some reference objects inside a plastic suitcase. The trajectory followed by the sensor has been planned considering the constraints previously mentioned to obtain unaliased high-resolution images of the total area of interest. Three reflecting spheres and two metal links are placed on top of absorbing material, creating a reflecting target area of about 40cm in width an 30cm in elevation. The obtained image is able to resolve the individual objects separated by about 5 cm (Fig. \ref{fig:SuitecaseResultsTop}).

\section{Conclusions and future work}

In this work we presented a concept for sensor-controlled trajectory planning for cognitive radar. As example, an UWB-Radar sensor was used in stripmap SAR mode for 3D-imaging. We showed the first results from an experiment with this sensor at 8GHz center frequency controlled by a 6-DOF manipulator arm. The test-objects located in a suitcase were resolved. The next step will be to investigate non-linear trajectories with additional constraints on the workspace of the manipulator arm to obtain volumetric representations of the detected objects.

\bibliography{IEEEabrv,UWBTrajectory}

\end{document}